\documentclass{article}

\PassOptionsToPackage{numbers,compress}{natbib}
\usepackage[final,dblblindworkshop]{neurips_2026}

\workshoptitle{Diffusion Language Models: Foundations, Efficiency, and Reasoning}

\usepackage[utf8]{inputenc}
\usepackage[T1]{fontenc}
\usepackage{hyperref}
\usepackage{url}
\usepackage{array}
\usepackage{booktabs}
\usepackage{makecell}
\usepackage{graphicx}
\usepackage{amsmath}
\usepackage{amsfonts}
\usepackage{bm}
\usepackage{xcolor}
\usepackage{tikz}
\usepackage{nicefrac}
\usepackage{microtype}
\usepackage{float}
\usepackage{wrapfig}
\usepackage{cleveref}

\newcolumntype{L}[1]{>{\raggedright\arraybackslash}m{#1}}
\newcolumntype{C}[1]{>{\centering\arraybackslash}m{#1}}

\title{Diff Diffusion: Predicting Edit Paths with \\ Diffusion Models for Efficient Code Editing}

\author{Xiaolong Xu\thanks{The first three authors contributed equally.}~~, Hao Jiang$^*$, Bo Jiang$^*$\thanks{Corresponding author.}~~, Zhiwen Deng, \\
\textbf{Ningyuan Sun, Bingzhou Chen, Jin Chen, Chenxing Wei}\\
ByteDance, Shenzhen, China\\
\texttt{\{xuxiaolong.06,jianghao.techxgenus,jiangbo.jacob,dengzhiwen.11,}\\
\texttt{sunningyuan,chenbingzhou,chenjin.723,weichenxing\}@bytedance.com}
}

\begin{document}

\maketitle

\begin{abstract}
Diffusion large language models (dLLMs) are appealing for code generation because they can update many tokens in parallel, yet many are optimized for sequence synthesis rather than edit prediction. This setting is poorly matched to modern AI-assisted IDEs, where edit prediction must infer the next code change from recent edits, the current file, and cursor-local context. Such changes are usually sparse and preserve most source tokens, so regenerating a full code segment wastes computation. We propose \textbf{Diff Diffusion}, a diffusion-inspired LLM framework for code editing that predicts edit paths from an original snippet to its revised form. Diff Diffusion represents denoising as structured edit operations, and combines this formulation with multi-stage training and efficient parallel decoding. By focusing computation on changed regions, Diff Diffusion supports accurate low-latency edits and multi-position parallel modification. Across public edit-completion benchmarks, edit-prediction evaluations, and online validation, Diff Diffusion matches or leads autoregressive baselines while delivering 2.78$\times$ faster inference on public edit-completion benchmarks and a 39.0\% P50 model-latency reduction in production. This work provides a new modeling perspective for AI applications that require high accuracy and low latency.
\end{abstract}

\begin{figure}[ht]
\centering
\includegraphics[width=0.7\linewidth]{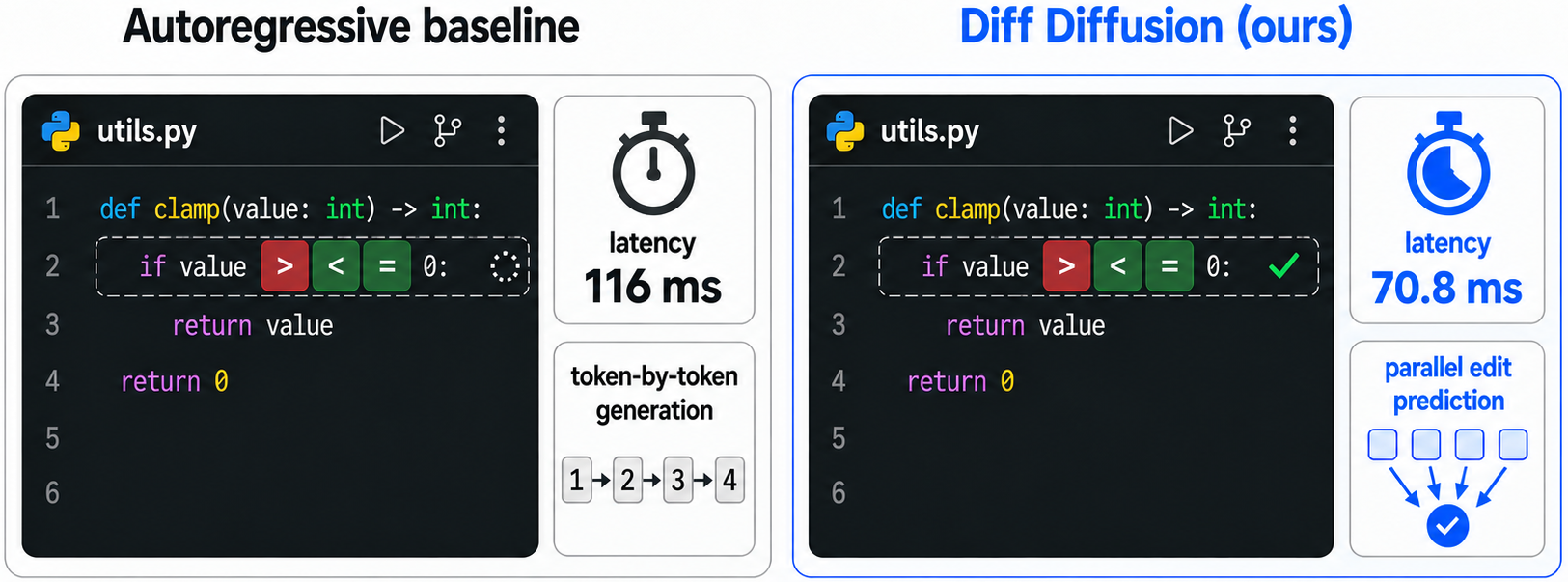}
\caption{Diff Diffusion models real-time IDE code edit prediction.}
\label{fig:ide_motivation}
\end{figure}

\section{Introduction}

Large language models (LLMs) have achieved remarkable progress in natural language understanding and generation, and have gradually extended this success to code intelligence tasks. Beyond the autoregressive (AR) paradigm, diffusion-based large language models (dLLMs) have become a promising alternative because they support parallel decoding and faster generation \cite{austin2021structured, li2025beyond}. By progressively denoising discrete noise tokens, dLLMs provide parallel generation capabilities and can make more efficient use of modern hardware computation. In terms of task performance, proprietary systems such as Mercury, Gemini Diffusion, and Seed Diffusion, as well as open-source projects such as DiffuCoder and Dream-Coder, have demonstrated competitive code-related capabilities \cite{labs2025mercury0, googledeepmind2025_gemini_diffusion, song2025seed, gong2025diffucoder0, xie2025dream0coder}.

However, current dLLMs mainly rely on masked diffusion to regenerate sequences from scratch, which does not match real programming workflows. In actual development, developers more often edit and refine existing code, and the changes are often small, local, and context-dependent. In IDE edit-prediction tasks, the system needs to predict one or more next code modifications in real time based on the user's edit history, the current file, and the rewrite range around the cursor. This latency-sensitive workflow is illustrated in \Cref{fig:ide_motivation}. Autoregressive models must serially output many unchanged fragments token by token in this task, cannot go back to correct already generated content, and are not well suited to processing multiple edits in parallel; masked diffusion often needs to denoise an entire sequence even when only a small change is required. Over the past year, a number of works on discrete diffusion and discrete flow matching have discussed token insertion, modification, and deletion \cite{song2025seed, wu2025fast0dllm0, li2025beyondfixed, zhang2025flexible, havasi2025editflows, kim2025anyorder, ding2026beyondmasks, wu2026dreamon, ren2026editbased}. However, they do not formulate edit operations as source-conditioned edit-path diffusion for latency-critical editing tasks, where unchanged source tokens should be preserved and skipped while only a few regions are revised.

Therefore, we propose \textbf{Diff Diffusion}, a diffusion-inspired LLM framework for code edit prediction. In addition to modeling variable-length generation with inserted and deleted tokens, Diff Diffusion builds edit-path prediction from the initial state and optimizes around low-latency inference. Unlike existing masked diffusion methods, Diff Diffusion explicitly models edit paths and predicts the sequence of edit operations required to transform an original code snippet into the revised code. By incorporating structured edit representations into the diffusion process, the framework avoids redundant regeneration and instead focuses on local updates. We also design a multi-stage training paradigm to progressively strengthen edit-prediction ability, and propose parallel decoding and inference-system optimization strategies to accelerate inference while maintaining accuracy. We first validate public edit-prediction ability on the C3PO open edit-prediction benchmark, then construct a dedicated edit-prediction evaluation benchmark to analyze improvements on Stay, Act, Multi-line, Roll-back, Delete, and other specific metrics, and finally validate the method in production.

In summary, our contributions are:
\begin{itemize}
    \item We propose \textbf{Diff Diffusion}, a source-conditioned diffusion-inspired framework that explicitly models and predicts code edit paths.
    \item We apply the framework to in-IDE edit prediction, using \texttt{same}, \texttt{delete}, \texttt{insert}, \texttt{replace}, and mask-to-token actions to reduce repeated decoding of unchanged code and support multi-position parallel modification.
    \item We design a multi-stage training paradigm and an efficient parallel decoding strategy to support accurate and fast inference.
    \item We evaluate the method on public benchmarks, edit-prediction evaluations, and production validation, demonstrating improvements in accuracy and latency.
\end{itemize}

\section{Diff Diffusion}

\begin{figure}[h]
\centering
\includegraphics[width=0.8\linewidth]{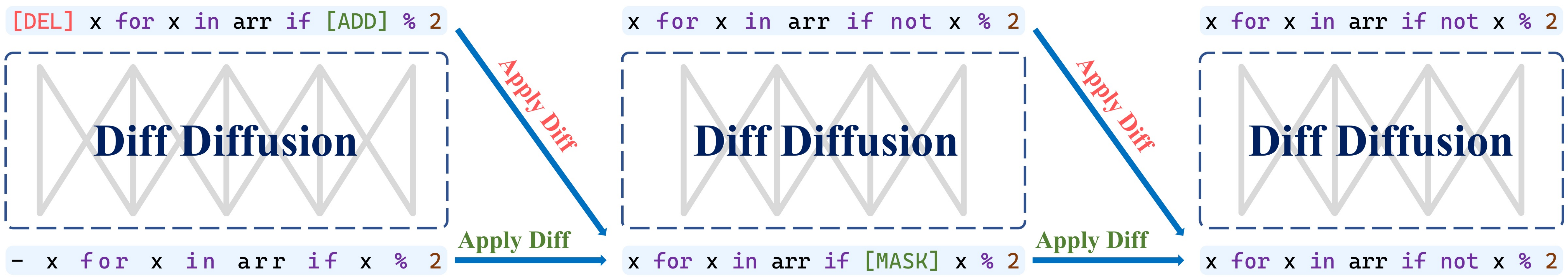}
\caption{Diff Diffusion framework.}
\label{fig:framework}
\end{figure}

Diff Diffusion is designed for edit prediction in IDEs, where the model must recommend local code rewrites in real time from the current file, cursor-local context, and recent editing history. We initialize Diff Diffusion from Dream-Coder \cite{xie2025dream0coder}, which provided the strongest diffusion backbone for code at the time we began our experiments, combining competitive performance on code-related tasks with favorable latency. We keep its Transformer architecture, but change the denoising target from reconstructing a full sequence to predicting an edit path. This section presents the diff-based diffusion process and the multi-stage learning strategy, the overall framework is shown in \Cref{fig:framework}. Appendix~\ref{app:related_work} provides additional context on diffusion language models and code editing.

\subsection{Diff-Based Diffusion Process}

Conventional masked diffusion randomly masks a subset of tokens and trains the model to recover them through iterative denoising. Edit prediction has a different objective: rather than recovering a full program fragment from a corrupted sequence, the model should transform an existing snippet into its revised form. We therefore define a diff-based diffusion process, in which the denoising trajectory is represented by edit operations. In addition to ordinary token prediction, the model predicts insertion, deletion, and replacement actions that explicitly define the edit path.

The denoising process from the source state to the target state can be viewed as diff prediction:
\begin{itemize}
    \item During training, the supervised target is the diff that specifies how the current state should be modified.
    \item During decoding, given $x_t$, the model predicts a diff $d_t$ such that applying it moves the sequence toward the previous denoising state, $x_{t-1}$.
\end{itemize}

\paragraph{Step-by-step formulation.}
\begin{enumerate}
    \item The initial state is the original code, denoted as $x_{\mathrm{src}} = [t_0, t_1, \ldots, t_n]$.
    \item The model assigns an edit label to each token in the current state:
    \begin{enumerate}
        \item keep the token unchanged by emitting its \texttt{token\_id};
        \item delete the token by emitting \texttt{del\_token\_id};
        \item insert a mask token before the current position by emitting \texttt{insert\_token\_id};
        \item replace the current token with a mask token by emitting \texttt{replace\_token\_id};
        \item fill a mask token by predicting a code \texttt{token\_id}.
    \end{enumerate}
    \item Tokens marked for deletion are removed, and the requested mask tokens are inserted or used as replacements.
    \item The process is repeated:
    \begin{enumerate}
        \item the model predicts code tokens for mask positions;
        \item the predicted tokens are again assigned edit labels;
        \item deleted tokens are removed, and newly inserted mask tokens are introduced.
    \end{enumerate}
\end{enumerate}

\paragraph{Advantages.}
This formulation retains the central benefit of masked diffusion: multiple positions can be updated in parallel. It also extends this benefit from generation to editing, allowing the model to apply edits at several locations in the same diffusion step. Because code edits usually preserve most source tokens and modify only a small region, the number of steps required to reach the target snippet is much smaller than in from-scratch masked diffusion or standard autoregressive generation. The insert and delete actions also make variable-length output a native part of the process, avoiding the fixed-canvas assumption of common masked-diffusion formulations.

\paragraph{Training Strategy.}
During training, we adjust the scheduler so that the initial state, target state, and intermediate diffusion states are sampled with different probabilities. This stabilizes learning across the trajectory. The intermediate states are constructed from the initial and target code through LCS alignment. Edit-prediction data is highly imbalanced, most tokens are unchanged, while insertion, deletion, and replacement labels are rare but determine the quality of the edit path. We therefore re-weight token types in the loss to increase the learning signal for edit actions.

\paragraph{Training Objective.}
Formally, let $x_t = [x_t^1, x_t^2, \ldots, x_t^n]$ denote the sequence at diffusion step $t$, and let $d_t = [d_t^1, d_t^2, \ldots, d_t^n]$ denote the corresponding diff labels. Each $d_t^i \in \mathcal{V}_{\mathrm{diff}}$, where $\mathcal{V}_{\mathrm{diff}}$ contains ordinary code-token ids as well as edit-control labels such as \texttt{delete}, \texttt{insert}, and \texttt{replace}. Re-emitting the current token id keeps a token unchanged, while predicting a token id at a mask position fills that mask. The model predicts a distribution $p_\theta(d_t^i \mid x_t)$ for each position.

The training objective is the weighted cross-entropy loss:
\[
\mathcal{L}(\theta) = - \sum_{t=1}^{T} \sum_{i=1}^{|x_t|} w(d_t^i) \cdot \log p_\theta(d_t^i \mid x_t),
\]
where $w(d_t^i)$ is a weight function that compensates for class imbalance:
\[
w(d_t^i) =
\begin{cases}
\lambda_{\text{edit}}, & d_t^i \in \{\texttt{insert}, \texttt{delete}, \texttt{replace}\}, \\
1, & \text{otherwise}.
\end{cases}
\]

At inference time, the model generates the revised snippet by iteratively sampling diffs and applying the update rule:
\[
x_{t-1} = \mathrm{ApplyDiff}(x_t, d_t),
\]
until the edited content no longer changes.

\subsection{Multi-Stage Diffusion Learning}

We train Diff Diffusion from a masked-diffusion backbone because the proposed formulation still requires mask-to-token prediction. Starting from this backbone transfers denoising ability to the edit setting, after which we progressively specialize the model for edit-path prediction.

\paragraph{Stage I: Large-Scale Git Commit Continuous Pretraining (CT).}
We first train the model on large-scale code-editing data so that it learns the diff-based modeling format and output space. This matches other evidence that dLLMs can surpass autoregressive models after large-scale training \cite{prabhudesai2025diffusionbeats, ni2025superdatalearners}. Commit diffs from open-source repositories provide natural supervision for source-to-target code transformations. In our setup, we use a multi-language subset of CommitPack \cite{DBLP:conf/iclr/MuennighoffLZZH24} as the starting point.

\paragraph{Stage II: Code-Editing SFT.}
After Stage I, the model has learned to predict differences along commit-level edit paths. Stage II adapts this ability to the edit-prediction input-output format used in interactive IDE scenarios. Each training example consists of the original snippet, surrounding context, historical edit information, and the target snippet. During training, we sample intermediate diffusion states on the edit path and supervise the model to predict the diff from the current state toward the target. By combining the C3PO training set \cite{brody2020structural} with high-quality edit-prediction data, Stage II converts general commit-level editing ability into a policy suitable for interactive edit prediction.

\paragraph{Stage III: Advanced Post-Training.}
The supervised objective in Stage II mainly learns the mapping from sampled intermediate states to the target state, whereas actual decoding repeatedly enters intermediate states produced by the model itself. Stage III therefore extends the training objective to learn more reliable edit directions. We apply this stage to a high-quality subset selected from the Stage-II SFT corpus. Inspired by discrete flow matching, we view the diff predicted from an intermediate diffusion state as a discrete modification direction from the current state toward the target state. This directional view supports our post-training schemes, Diff-DPO and AGRPO, both defined over denoising transitions. Let $c$ denote the IDE context, including the source snippet, surrounding code, and recent edit history, and let $y$ denote the target snippet. For a transition $x_t \rightarrow x_{t-1}$ induced by a realized diff $d_t$, we define the transition log-probability as
\[
\ell_\theta(x_{t-1}\mid x_t,c)
= \sum_{i=1}^{|d_t|} \log p_\theta(d_t^i \mid x_t,c),
\qquad
x_{t-1}=\mathrm{ApplyDiff}(x_t,d_t).
\]
This transition-level view lets us apply preference learning and reinforcement learning directly to edit directions rather than to left-to-right token emissions. Diff-DPO uses paired final states, where the positive state $y^{+}$ is the ground-truth target snippet and the negative state $y^{-}$ is an erroneous edit sampled from the model. During policy decoding, we sample an intermediate state $x_t$; from this same state, we construct $x_{t-1}^{+}$ and $x_{t-1}^{-}$ as the next denoising states along shortest edit paths to $y^{+}$ and $y^{-}$, respectively. With a reference policy $\pi_{\mathrm{ref}}$, we optimize a Diff Diffusion version of DPO \cite{rafailov2023direct, wallace2023diffusion}:
\[
\mathcal{L}_{\mathrm{Diff\mbox{-}DPO}}(\theta)
= -\mathbb{E}_{(c,x_t,x_{t-1}^{+},x_{t-1}^{-})}
\left[\log\sigma\left(\beta(\Delta_\theta-\Delta_{\mathrm{ref}})\right)\right],
\]
where $\Delta_\pi=\ell_\pi(x_{t-1}^{+}\mid x_t,c)-\ell_\pi(x_{t-1}^{-}\mid x_t,c)$ for $\pi\in\{\pi_\theta,\pi_{\mathrm{ref}}\}$. This loss increases the model's preference for the ground-truth edit direction over the erroneous model-generated direction, while the reference gap keeps the update anchored to the Stage-II policy.

For online reinforcement learning, we survey recent RL methods for diffusion and flow-matching models and adopt AGRPO because it gives the strongest performance with a simple implementation in our attempts \cite{shao2024deepseekmath, black2023training, wang2025revolutionizing, liu2025flowgrpo, wei2026lfpo, zhan2025agrpo}. For each example $q=(c,y)$, the old policy samples a group of $G$ complete edit trajectories $\{\tau_i\}_{i=1}^{G}$, where $\tau_i=\{(x_t^i,d_t^i,x_{t-1}^i)\}_{t=1}^{m_i}$ and the final state is $\hat{y}_i$. We use a terminal reward with a repetition penalty,
\[
R_i = \mathbf{1}\{\hat{y}_i=y\} - \lambda_{\mathrm{rep}}\mathrm{Rep}(\hat{y}_i),
\qquad
A_i=\frac{R_i-\frac{1}{G}\sum_{j=1}^{G}R_j}
{\mathrm{std}(\{R_j\}_{j=1}^{G})+\delta},
\]
which converts absolute rewards into group-relative advantages. We view diffusion generation as a multi-step Markov denoising process: instead of summing autoregressive output-token log-probabilities, AGRPO optimizes the policy objective over denoising transitions. For a sampled subset of diffusion steps $\mathcal{S}_i \subseteq \{1,\ldots,m_i\}$, we maximize
\[
\begin{aligned}
\mathcal{J}_{\mathrm{AGRPO}}(\theta)
= {} & \mathbb{E}_{q,\,\{\tau_i\}\sim\pi_{\mathrm{old}}}
\Biggl[
\frac{1}{G}\sum_{i=1}^{G}\frac{1}{|\mathcal{S}_i|}
\sum_{t\in\mathcal{S}_i}
\biggl(
\min\!\left\{r_{i,t}(\theta)A_i,
\operatorname{clip}(r_{i,t}(\theta),1-\epsilon,1+\epsilon)A_i\right\}
\\
&\qquad\qquad
-\beta_{\mathrm{KL}}
D_{\mathrm{KL}}\!\left(
\pi_\theta(\cdot\mid x_t^i,c)\,\|\,\pi_{\mathrm{ref}}(\cdot\mid x_t^i,c)
\right)
\biggr)
\Biggr],
\end{aligned}
\]
where $r_{i,t}(\theta)=\exp(\ell_\theta(x_{t-1}^i\mid x_t^i,c)-\ell_{\mathrm{old}}(x_{t-1}^i\mid x_t^i,c))$. Sampling only a small number of diffusion steps gives a low-cost Monte Carlo estimate of the expectation over diffusion steps, while retaining the trajectory-level reward that reflects whether the final edit is correct and optimizing complete edit trajectories with controllable compute cost.

\section{Inference and Infrastructure}
To support high-volume online edit-prediction requests, we build a dedicated inference engine for Diff Diffusion. Its system design follows recent work on accelerating diffusion language models through block-wise parallel decoding, approximate KV-Cache reuse, and system-level inference optimization \cite{song2025seed, wu2025fast0dllm0, ma2025dinfer}. In contrast to general-purpose generation settings, our target is an IDE code-editing task, where the inference engine must reduce the number of edit iterations and translate this reduction into end-to-end latency gains.

\subsection{Block-wise Parallel Edit Decoding}
AR models must generate the complete target snippet sequentially, whereas Diff Diffusion treats the current code snippet as an editable state. At each diffusion step, the model predicts \texttt{same}, \texttt{delete}, \texttt{insert}, \texttt{replace}, and mask-to-token actions in parallel. Non-mask positions are restricted to remain unchanged or special edit actions, so replacements must go through \texttt{replace} and subsequent mask filling. Block-wise decoding further partitions the rewrite range into local blocks, applies high-confidence edits within each block in parallel, and uses unchanged tokens from the original code as left and right anchors to constrain edit boundaries. In addition, insertion length is adaptive: when the inference engine assigns high confidence to an insertion region, a single insert action can emit several consecutive tokens, preventing long-completion cases from degenerating into token-by-token insertion.

\subsection{Approximate KV-Cache in Editing Scenarios}
We use approximate KV-Cache to reuse bidirectional cache during denoising. Diff Diffusion uses two levels of cache optimization. The first level targets stable context within a service request. Edit-prediction requests contain a system prompt, the current file, and historical edit information, so our inference engine reuses KV states for stable prefixes and converged blocks. The second level targets suffix reuse in long-sequence editing. When only a middle region needs to be modified, a large code suffix after the edit point remains unchanged. We therefore follow the idea of PIE and reuse suffix KV states after RoPE position correction \cite{he2024let}. This strategy is designed for local code edits with long unchanged suffixes, and in our tests it preserves accuracy while providing substantial end-to-end latency gains.

\section{Experiments}
We evaluate Diff Diffusion around the edit-prediction setting targeted by the method. The experiments cover four questions: whether the model learns public edit-completion patterns, whether post-training improves IDE-specific edit-prediction metrics, whether parallel edit decoding reduces inference latency, and whether the latency gains transfer to production traffic. Large-scale edit pretraining and supervised fine-tuning are run on an Ascend 910B cluster; reinforcement-learning training and evaluation are run on NVIDIA H20 GPUs. Our autoregressive baseline is the production edit-prediction model used before the diffusion cutover, trained from Seed-Coder \cite{seed2025seed0coder0}. It receives the same recent-edit history, file context, and edit range as Diff Diffusion, and generates the revised snippet from left to right. In the IDE edit-prediction evaluation, we also compare with Mercury-Coder, a diffusion-based model from Mercury \cite{labs2025mercury0}, which provides an edit-prediction endpoint. Detailed training and inference settings are provided in Appendix~\ref{app:hyperparameters}.

\subsection{Data and Evaluation Setup}
The training and evaluation data are summarized in Appendix~\ref{app:data_statistics}. CommitPack provides large-scale commit-level supervision for Stage I pretraining, the edit-prediction training data supports Stage II and Stage III post-training, C3PO provides an open edit-completion benchmark, and our edit-prediction evaluation set measures the IDE behaviors that matter for next-edit prediction.

\subsection{Open Edit-Completion Benchmark}
C3PO Edit Completion \cite{brody2020structural} asks a model to predict the remaining edit from contextual edits that have already occurred. This benchmark directly tests whether the model captures structured edit patterns rather than only token-level continuation. We report result accuracy, average decoding steps per sample, and average inference time per sample.

\begin{table}[h]
\centering
\caption{Performance comparison on the C3PO edit-completion task.}
\label{tab:c3po_results}
\begingroup
\setlength{\tabcolsep}{2pt}
\renewcommand{\arraystretch}{1.05}
\begin{tabular}{@{}L{0.19\linewidth}C{0.15\linewidth}C{0.14\linewidth}C{0.22\linewidth}C{0.17\linewidth}@{}}
\toprule
\textbf{Model} &
\shortstack{\textbf{Result}\\\textbf{Accuracy (\%)}} &
\shortstack{\textbf{Steps}\\\textbf{per Sample}} &
\shortstack{\textbf{Inference Time}\\\textbf{per Sample (ms)}} &
\shortstack{\textbf{Speedup}\\\textbf{Relative to AR}} \\
\midrule
AR & \underline{83.18} & 14.62 & 403 & $-$ \\
Diff Diffusion & 81.33 & \textbf{3.90} & \textbf{145} & \textbf{$2.78\times$} \\
\bottomrule
\end{tabular}
\endgroup
\end{table}

\Cref{tab:c3po_results} shows that Diff Diffusion reaches 81.33\% result accuracy, close to the AR baseline, while reducing the average number of decoding steps from 14.62 to 3.90. This accuracy gap is consistent with prior findings that dLLMs benefit from larger-scale training and more optimization epochs before surpassing AR models \cite{prabhudesai2025diffusionbeats, ni2025superdatalearners}, while the C3PO training set is relatively small. The step reduction translates into 145 ms inference time per sample, compared with 403 ms for AR, yielding a $2.78\times$ speedup on the public edit-completion task.

\subsection{IDE Edit-Prediction Evaluation}
We further evaluate on an edit-prediction set that mirrors next-edit prediction in IDEs. The examples are collected from diverse editing scenarios. Each example contains recent user edits, the current file, and an edit range; the model must decide whether to recommend an edit and, when appropriate, generate the local code transformation. The evaluation set is organized into five subsets, each targeting a specific IDE behavior:
\begin{itemize}
    \item \textbf{Act}: valid rewrite scenarios where recommendations should be triggered; this metric measures the proportion of correct local code transformations and reflects the model's ability to identify and complete the next edit.
    \item \textbf{Stay}: quiet scenarios where no recommendation is needed; this metric measures whether the model keeps the original code unchanged and reflects its ability to suppress unnecessary suggestions.
    \item \textbf{Multi-line}: cross-line synchronized rewrite scenarios, such as consecutive changes to function parameters, call chains, or repeated patterns; this metric measures the model's ability to complete multi-line edits.
    \item \textbf{Roll-back}: cases where the user has just made a local code change; this metric measures whether the model incorrectly reverts that recent modification and reflects its ability to preserve edit intent and history.
    \item \textbf{Delete}: local rewrite scenarios where surrounding context should be preserved; this metric measures unreasonable deletion of valid code and reflects the model's ability to control edit boundaries.
\end{itemize}

\begin{table}[h]
\centering
\caption{Results on the IDE edit-prediction evaluation set.}
\label{tab:edit_prediction_results}
\begingroup
\setlength{\tabcolsep}{2pt}
\renewcommand{\arraystretch}{1.15}
\begin{tabular}{@{}L{0.44\linewidth}C{0.08\linewidth}C{0.08\linewidth}C{0.13\linewidth}C{0.12\linewidth}C{0.08\linewidth}@{}}
\toprule
\textbf{Model} & \textbf{\makecell[c]{Act\\(\%) $\uparrow$}} & \textbf{\makecell[c]{Stay\\(\%) $\uparrow$}} & \textbf{\makecell[c]{Multi-line\\(\%) $\uparrow$}} & \textbf{\makecell[c]{Roll-back\\(\%) $\downarrow$}} & \textbf{\makecell[c]{Delete\\(\%) $\downarrow$}} \\
\midrule
AR & 45.90 & 91.73 & 68.42 & 9.0 & 1.0 \\
Mercury-Coder & 31.97 & 54.32 & 36.84 & 52.5 & 19.5 \\
Diff Diffusion + CT & 41.80 & 89.21 & 68.42 & \textbf{7.0} & \textbf{0.5} \\
Diff Diffusion + CT + SFT & 49.18 & 89.21 & \textbf{73.68} & 9.5 & \textbf{0.5} \\
Diff Diffusion + CT + SFT + DPO & 49.18 & 92.45 & 68.42 & 7.5 & \textbf{0.5} \\
Diff Diffusion + CT + SFT + AGRPO & \textbf{50.00} & \textbf{96.40} & \textbf{73.68} & 10.0 & \textbf{0.5} \\
\bottomrule
\end{tabular}
\endgroup
\end{table}

As shown in \Cref{tab:edit_prediction_results}, the base Diff Diffusion model after CT substantially improves over Mercury-Coder on all five IDE edit-prediction metrics and already matches AR on Multi-line while reducing both Roll-back and Delete, reflecting the benefit of modeling edits as sparse transformations over mostly preserved code. Post-training then strengthens recommendation quality. SFT raises Act to 49.18\% and Multi-line to 73.68\%, DPO improves Stay while keeping Roll-back low, and AGRPO obtains the strongest Act and Stay scores, reaching 50.00\% Act and 96.40\% Stay while keeping Delete at 0.5\%.

\subsection{Inference Latency}
We measure inference latency on the edit-prediction evaluation set. Before discussing the results, we first define the metrics reported in \Cref{tab:latency_results}. Edit Range is the token span that the model may rewrite, Output Length is the average number of output tokens, Decoding Steps is the average number of decoding iterations needed for one edit, and Total Time is the average end-to-end time per sample.

\begin{table}[h]
\centering
\caption{Inference speed on the edit-prediction evaluation set.}
\label{tab:latency_results}
\begingroup
\setlength{\tabcolsep}{2pt}
\renewcommand{\arraystretch}{1.16}
\begin{tabular}{@{}L{0.26\linewidth}C{0.17\linewidth}C{0.16\linewidth}C{0.16\linewidth}C{0.17\linewidth}@{}}
\toprule
\textbf{Method} & \textbf{\makecell[c]{Edit Range\\(tokens)}} & \textbf{\makecell[c]{Output\\Length}} & \textbf{\makecell[c]{Decoding\\Steps}} & \textbf{\makecell[c]{Total Time\\(ms)}} \\
\midrule
AR & 140 & 126 & 126 & 3964 \\
Diff Diffusion & 140 & 110 & \textbf{4.90} & \textbf{415} \\
Diff Diffusion + PIE & 140 & 110 & 5.00 & 420 \\
Diff Diffusion & 1024 & 1008 & \textbf{7.56} & 784 \\
Diff Diffusion + PIE & 1024 & 1012 & 8.78 & \textbf{676} \\
\bottomrule
\end{tabular}
\endgroup
\end{table}

For a 140-token rewrite range, diffusion decoding compresses generation from 126 AR steps to about five edit steps, reducing total time from 3964 ms to about 415--420 ms. PIE does not help this short case because both variants are memory-bound at this scale, and the extra suffix-cache bookkeeping and position correction slightly outweigh the saved computation. On the 1024-token long-rewrite setting, approximate suffix-cache reuse through PIE further lowers total time from 784 ms to 676 ms. These results show that the main efficiency gain comes from few-step parallel editing, while approximate suffix-cache reuse provides additional end-to-end savings when the unchanged suffix is long.

\subsection{Online Validation}
\label{sec:online_ab}
Finally, we validate Diff Diffusion in a real IDE edit-prediction scenario with online A/B testing. Appendix~\ref{app:ab_setup} describes the relevant details. The online metrics used in this environment are as follows \cite{ziegler2022productivity,maddila2025aiassisted}:
\begin{itemize}
    \item \textbf{Acceptance rate}: the proportion of valid displayed edit suggestions that are accepted by users. Suggestions rejected after too short a display time and other invalid impressions are filtered out.
    \item \textbf{CPO (Characters per Opportunity)}: the number of characters contributed by AI after each user input. We compute CPO from the edit distance before and after accepting an edit suggestion, measuring the effective rewrite amount produced by one user action.
    \item \textbf{Accepted lines}: the number of accepted code lines per user within the evaluation window, measuring the practical line-level code benefit obtained by users.
    \item \textbf{P50 model inference latency}: the median model inference latency for generating the model suggestion.
\end{itemize}
Both AR and Diff Diffusion are served through dedicated inference engines with tensor parallel size 4 and FP8 quantization, while the AR model additionally enables NGram speculative sampling to reduce latency. \Cref{tab:online_results} reports the production edit-prediction validation results.

\begin{table}[h]
\centering
\caption{Production A/B results for IDE edit prediction.}
\label{tab:online_results}
\begingroup
\setlength{\tabcolsep}{2pt}
\renewcommand{\arraystretch}{1.16}
\begin{tabular}{@{}L{0.21\linewidth}C{0.17\linewidth}C{0.12\linewidth}C{0.17\linewidth}C{0.22\linewidth}@{}}
\toprule
\textbf{Model} & \textbf{\makecell[c]{Acceptance\\rate}} & \textbf{CPO} & \textbf{\makecell[c]{Accepted\\lines}} & \textbf{\makecell[c]{P50 model\\latency}} \\
\midrule
AR & 23.8\% & \underline{1.06} & 93.6 & 116 ms \\
Diff Diffusion & \textbf{24.0\%} & 1.05 & \textbf{102.6} & \textbf{70.8 ms} \\
\bottomrule
\end{tabular}
\endgroup
\end{table}

\begin{wrapfigure}[11]{r}{0.45\linewidth}
\centering
\resizebox{\linewidth}{!}{\begingroup
\sffamily
\small
\begin{tikzpicture}[x=1cm,y=0.010cm]
  \definecolor{ddblue}{RGB}{33,113,181}
\definecolor{ddar}{RGB}{70,70,70}
\definecolor{ddgrid}{RGB}{226,226,226}
\definecolor{ddaxis}{RGB}{40,40,40}

\foreach \y/\label in {0/0,50/50,100/100,150/150,200/200,250/250,300/300} {
  \draw[ddgrid] (0,\y) -- (7.55,\y);
  \draw[ddaxis] (-0.04,\y) -- (0,\y);
  \node[anchor=east,font=\scriptsize] at (-0.12,\y) {\label};
}

\draw[->,ddaxis] (0,0) -- (7.85,0);
\draw[->,ddaxis] (0,0) -- (0,318);
\node[anchor=north] at (3.78,-44) {Edit distance};
\node[anchor=south,rotate=90] at (-0.82,150) {Latency (ms)};

\foreach \x/\label in {0/{0--10},1.2/{11--20},2.4/{21--30},3.6/{31--60},4.8/{61--90},6.0/{91--120},7.2/{$>$120}} {
  \draw[ddaxis] (\x,0) -- (\x,-6);
  \node[anchor=north,font=\scriptsize] at (\x,-12) {\label};
}

\draw[very thick,ddar]
  (0,78) -- (1.2,93) -- (2.4,104) -- (3.6,121) --
  (4.8,139) -- (6.0,173) -- (7.2,213);
\foreach \x/\y in {0/78,1.2/93,2.4/104,3.6/121,4.8/139,6.0/173,7.2/213}
  \fill[ddar] (\x,\y) circle (1.5pt);

\draw[very thick,ddblue]
  (0,31) -- (1.2,49) -- (2.4,66) -- (3.6,84) --
  (4.8,104) -- (6.0,135) -- (7.2,197);
\foreach \x/\y in {0/31,1.2/49,2.4/66,3.6/84,4.8/104,6.0/135,7.2/197}
  \fill[ddblue] (\x,\y) circle (1.5pt);

\draw[very thick,ddar] (4.28,286) -- (4.58,286);
\fill[ddar] (4.43,286) circle (1.5pt);
\node[anchor=west,font=\scriptsize] at (4.66,286) {AR};
\draw[very thick,ddblue] (5.18,286) -- (5.48,286);
\fill[ddblue] (5.33,286) circle (1.5pt);
\node[anchor=west,font=\scriptsize] at (5.56,286) {Diff Diffusion};
\end{tikzpicture}
\endgroup}
\caption{Latency by edit-distance bucket.}
\label{fig:edit_distance_latency}
\end{wrapfigure}
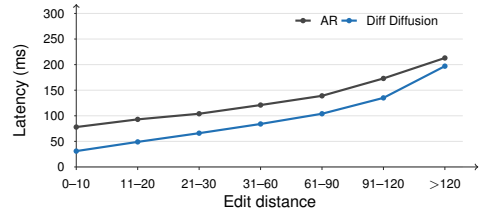

The online results confirm that the local decoding advantage transfers to the production serving path. Diff Diffusion keeps acceptance rate and CPO comparable to AR, increases accepted lines per user by 9.6\%, and reduces P50 model inference latency by 39.0\%. The edit-distance breakdown in \Cref{fig:edit_distance_latency} further shows that the latency advantage holds across different edit sizes, establishing the practical value of edit-path diffusion for interactive coding systems where recommendation quality and response time must be optimized together.

\section{Discussion}
This paper presents Diff Diffusion, a diffusion-inspired framework that explicitly models code edit paths. Our study shows that aligning diffusion models with the intrinsic properties of edit prediction, namely locality, incrementality, and context dependence, provides clear advantages over traditional autoregressive methods that regenerate the entire sequence. For edit prediction, these advantages are concentrated in skipping unchanged tokens, editing multiple positions in parallel, and reducing inference latency through approximate KV-Cache reuse, PIE, and multi-token insertion. Our current limitations are that we have not yet developed a deeper theoretical analysis of edit-path diffusion, and our evaluation focuses on latency-critical IDE edit prediction. In the future, we will extend the framework to broader editing settings, including instruction-guided editing, repository-level changes, and agent workflows that combine planning, tool use, and human review.

\begin{ack}
This work was conducted at ByteDance and received no third-party funding or support. The authors declare no competing interests beyond their disclosed affiliation with ByteDance.
\end{ack}

\bibliographystyle{plainnat}
\bibliography{references}

\appendix
\section{Related Work}
\label{app:related_work}

\paragraph{Diffusion Language Models (dLLMs).}
Diffusion models have been widely adopted in continuous domains such as image generation \cite{rombach2021high0resolution}. Recently, discrete diffusion models have emerged as an alternative paradigm for large language models because their denoising process can update multiple positions in parallel rather than following a strict left-to-right order \cite{austin2021structured}. Masked diffusion is the dominant formulation, LLaDA and Dream scale this approach to 7B--8B parameters with performance comparable to autoregressive models, and subsequent LLaDA-family work extends it to preference alignment, long-context modeling, multimodal instruction tuning, sparse MoE, and 100B-scale training \cite{nie2025large, ye2025dream, zhu2025llada15, liu2025longllada, you2025lladav, zhu2025lladamoe, bie2025llada20, bie2026llada21, nie2026improved}. More recent work further studies variable-length denoising, insertion/deletion operations, edit-based refinement, and code infilling \cite{wu2025fast0dllm0, ma2025dinfer, li2025beyondfixed, zhang2025flexible, havasi2025editflows, kim2025anyorder, ding2026beyondmasks, inclusionai2026llada22, wu2026dreamon, ren2026editbased}. These works show that insertion and deletion operations have become an important direction in dLLM modeling. Diff Diffusion focuses on edit-path modeling from the original snippet to the target snippet, as well as the system-level benefits of this modeling for low-latency tasks. The specific differences are shown in \Cref{tab:dllm_edit_ops}.

\begin{table}[h]
\centering
\caption{Representative dLLMs with edit operations.}
\label{tab:dllm_edit_ops}
\begingroup
\setlength{\tabcolsep}{2pt}
\renewcommand{\arraystretch}{1.20}
\begin{tabular}{@{}L{0.16\linewidth}C{0.14\linewidth}C{0.24\linewidth}C{0.11\linewidth}C{0.16\linewidth}C{0.12\linewidth}@{}}
\toprule
\textbf{Method} &
\shortstack{\textbf{Initial}\\\textbf{State}} &
\shortstack{\textbf{Edit}\\\textbf{Actions}} &
\shortstack{\textbf{Edit Task}\\\textbf{Focused}} &
\shortstack{\textbf{Main}\\\textbf{Goal}} &
\shortstack{\textbf{Production}\\\textbf{Grade}} \\
\midrule
Seed Diffusion \cite{song2025seed} &
Masked sequence &
Mask-to-token &
Partial &
Quality + latency &
No \\
LLaDA2.1 \cite{bie2026llada21} &
Masked sequence &
\makecell{Mask-to-token,\\token replacement} &
No &
Latency + quality &
No \\
LLaDA2.2 \cite{inclusionai2026llada22} &
Masked sequence &
\makecell{Keep, substitute,\\delete, insert} &
Partial &
Self-correction &
No \\
Edit Flows \cite{havasi2025editflows} &
Random sequence &
Insert, delete, substitute &
No &
Variable length &
No \\
DID \cite{ding2026beyondmasks} &
Masked sequence &
Delete, insert &
No &
Variable length &
No \\
DreamOn \cite{wu2026dreamon} &
Masked span &
Expand, delete, fill mask &
Partial &
\makecell{Variable-length\\infilling} &
No \\
\textbf{Diff Diffusion} &
\textbf{Source span} &
\makecell{\textbf{Same, delete, insert,}\\\textbf{replace, fill mask}} &
\textbf{Yes} &
\textbf{Edit latency + quality} &
\textbf{Yes} \\
\bottomrule
\end{tabular}
\endgroup
\end{table}

\paragraph{Code LLMs.}
LLMs have become a central model family for code intelligence. Proprietary models such as Claude, GPT, and Gemini exhibit strong coding capabilities, while open-source models including Qwen2.5-Coder and Seed-Coder advance code modeling through pretraining on large-scale code corpora, instruction tuning with self-supervised data, commit histories, and evolutionary instructions, and alignment methods such as DPO and GRPO \cite{hui2024qwen25coder, seed2025seed0coder0}. Diffusion-based code models are now emerging alongside these autoregressive systems: proprietary models such as Mercury, Gemini Diffusion, and Seed Diffusion, as well as open-source efforts such as DiffuCoder and Dream-Coder, have shown competitive performance on code-related tasks \cite{labs2025mercury0, googledeepmind2025_gemini_diffusion, song2025seed, gong2025diffucoder0, xie2025dream0coder}.

\paragraph{Code Editing.}
Code editing is an important task in AI-assisted programming, extending the classical program synthesis paradigm to enable fine-grained code modifications. It is widely integrated into popular tools such as Continue, Aider, Copilot, Cursor, and Trae and supports well-known next-edit prediction features, including Cursor Tab and Copilot NES \cite{vscode2026copilotnes, cursor2026tab}. Code editing leverages various forms of software engineering knowledge, including historical edit records, user instructions, and compiler error messages. Numerous benchmarks have been proposed to evaluate model performance. Function-level and file-level benchmarks include C3PO, APEval, CanItEdit, CodeEditorBench, Aider, and HumanEvalFix \cite{brody2020structural, jiang2024cursorcore, cassano2023edit, guo2024codeeditorbench, aider, DBLP:conf/iclr/MuennighoffLZZH24}, while repository-level evaluations include CoEditor and SWE-bench \cite{wei2023coeditor, DBLP:conf/iclr/JimenezYWYPPN24}.

\section{Data Statistics}
\label{app:data_statistics}

\begin{table}[h]
\centering
\caption{Training and evaluation data.}
\label{tab:datasets}
\begingroup
\setlength{\tabcolsep}{3pt}
\renewcommand{\arraystretch}{1.24}
\begin{tabular}{@{}L{0.15\linewidth}L{0.19\linewidth}L{0.34\linewidth}L{0.26\linewidth}@{}}
\toprule
\textbf{Dataset / Benchmark} & \textbf{Role} & \textbf{Scope} & \textbf{Scale} \\
\midrule
CommitPack\newline subset & Stage I edit-path\newline pretraining & Git commit diffs in Python, C++, Java, JavaScript, Go, and Rust & 87B tokens \\
Edit prediction\newline training data & Stage II / III IDE\newline post-training & High-quality next-edit examples collected from multiple scenarios & 30M examples \\
C3PO Edit Completion & Public benchmark and SFT data & C\# code edit examples from 53 GitHub projects & 39K train / 5K eval examples \\
Edit prediction\newline evaluation set & In-house IDE\newline benchmark & Act, Stay, Multi-line, Roll-back,\newline and Delete cases & 122 / 278 / 19 / 200 / 200\newline examples each subset \\
\bottomrule
\end{tabular}
\endgroup
\end{table}

\paragraph{Asset Licenses and Terms.}
We use existing assets under their stated licenses or service terms. CommitPack is distributed under the MIT License \cite{DBLP:conf/iclr/MuennighoffLZZH24}. C3PO's public code and data release is under the MIT License, with examples drawn from the C\# GitHub projects listed by the benchmark authors \cite{brody2020structural}. The Dream-Coder initialization checkpoint and companion code are released under the Apache-2.0 License \cite{xie2025dream0coder}. The AR baseline is trained from Seed-Coder, whose public release is under the MIT License \cite{seed2025seed0coder0}. Mercury-Coder is used through Inception Labs' hosted API under its service terms \cite{labs2025mercury0}.

\section{Training and Inference Hyperparameters}
\label{app:hyperparameters}

In our CT and SFT experiments, we use weighted cross-entropy over edit labels: ordinary token and \texttt{same} predictions use weight 1, while the rarer structural actions \texttt{insert}, \texttt{delete}, and \texttt{replace} use weight 5. Prompt tokens are excluded from the prediction loss. All training stages use the AdamW optimizer, gradient clipping at 1.0, and a warmup-decay learning-rate schedule. For advanced post-training, we first run a reference-only Diff-DPO pass to cache the chosen and rejected transition log-probabilities, then optimize the actor with the Diff-DPO loss using $\beta=0.1$. AGRPO samples $G=8$ complete edit trajectories for each prompt. When a trajectory contains more than 25 denoising steps, 25 transitions are sampled for the policy-gradient estimate, while shorter trajectories optimize all transitions. In our inference settings, the context window is set to 5.5K tokens, the maximum generation length is set to 356 tokens, and the maximum number of iterations is set to 178.

\section{Online A/B Experiment Setup}
\label{app:ab_setup}

We conduct an online A/B experiment within an existing IDE edit-prediction feature. Eligible users are randomly assigned to persistent control and treatment groups, which receive the AR baseline and Diff Diffusion, respectively, under the same user-facing configuration. We apply consistent eligibility and validity criteria to both groups and compute aggregate metrics over valid suggestion impressions and successful requests during the evaluation window.

\end{document}